\pdfoutput=1

\documentclass[a4paper]{jpconf}
\usepackage{graphicx}
\usepackage{amssymb}

\begin{document}

\title{Towards a liquid Argon TPC without evacuation: filling 
of a 6 m$^3$ vessel with argon gas from air to ppm impurities concentration through flushing}

\author{A. Curioni$^1$, L. Epprecht$^1$, A. Gendotti$^1$,
L. Knecht$^1$, D. Lussi$^1$, A.~Marchionni$^1$, G. Natterer$^1$, F. Resnati$^1$, A. 
Rubbia$^1$, J. Coleman$^2$, M.~Lewis$^2$, K. Mavrokoridis$^2$, K. McCormick$^2$, C. Touramanis$^2$}

\address{$^1$ ETH Zurich, 101 Raemistrasse, CH-8092 Zurich, Switzerland}
\address{$^2$ University of Liverpool, Liverpool, United Kingdom}

\ead{alessandro.curioni@cern.ch}

\begin{abstract}
In this paper we present a successful experimental test of filling a volume of 6 m$^3$ with argon gas,
starting from normal ambient air and reducing the impurities content down to few parts per million (ppm) oxygen equivalent. 
This level of contamination was directly monitored measuring the slow component of the scintillation light 
of the Ar gas, which is sensitive to {\it all} sources of impurities affecting directly the argon scintillation.
\end{abstract}


\section{Introduction}

GLACIER (Giant Liquid Argon Charge Imaging ExpeRiment) is a proposed very large LAr TPC with a well defined 
conceptual design~\cite{Rubbia:2004tz,Rubbia:2009md}.
The proposed cryostat is a single module cryo-tank based on industrial liquefied natural gas (LNG) 
technology, with a cylindrical shape giving an optimal surface-to-volume ratio. The detector design is meant to be highly 
scalable, up to a total mass of at least 100 kton. 

One severe key technical challenge of the GLACIER design is to achieve long drift paths of free ionization electrons 
in liquid argon over a very long distance, possibly as large as  20 meters. The GLACIER concept therefore
relies on a double phase operation with a readout
with charge amplification to compensate for the S/N loss due to diffusion and remaining charge attenuation from
impurity attachment during the long drift path, 
but in addition requires very special care to achieve and maintain very high purity in the liquid argon,
at the level of $\lesssim 20~$ppt oxygen equivalent~\cite{ARubbia}, and a very high voltage to create the drift field to 
quickly drift the electron cloud across the liquid argon volume~\cite{Horikawa}. 

Argon purification systems based on oxygen reactants and molecular sieves, commercially available, have been studied 
since many years as part of the R\&D for liquid Argon TPCs~\cite{Bettini:1990rj,Benetti:1993mf}, and are known to be very effective 
in reducing the oxygen contamination and also water, carbon oxydes, fluorates elements or hydrocarbons to very small levels.
During the ICARUS 50L exposure to the WANF neutrino beam at CERN~\cite{Arneodo:2006ug}, a free electron lifetime of 15~ms, 
corresponding to 20~ppt oxygen equivalent, was achieved using such standard (gas) purification methods.

However, like the ICARUS 50L, all LAr TPCs 
operated up to now were housed in a vacuum vessel, and ultra-high vacuum evacuation 
was always performed prior to cooling and filling with liquid argon. This will not be possible
for a LNG tank, because this kind of vessel, although a time tested solution,
cannot withstand under- or over-pressures above 50-100~mbar, so evacuation is impossible. As of today, the experimental
demonstration that high liquid argon purity can be achieved and maintained over long periods
of operation in an LNG tank, is lacking. As a first step towards this milestone,
we started studying new methods and scalable techniques to manage non-evacuable vessels.

\begin{figure}[tb]
   \centering	
   \includegraphics[width=0.75\textwidth]{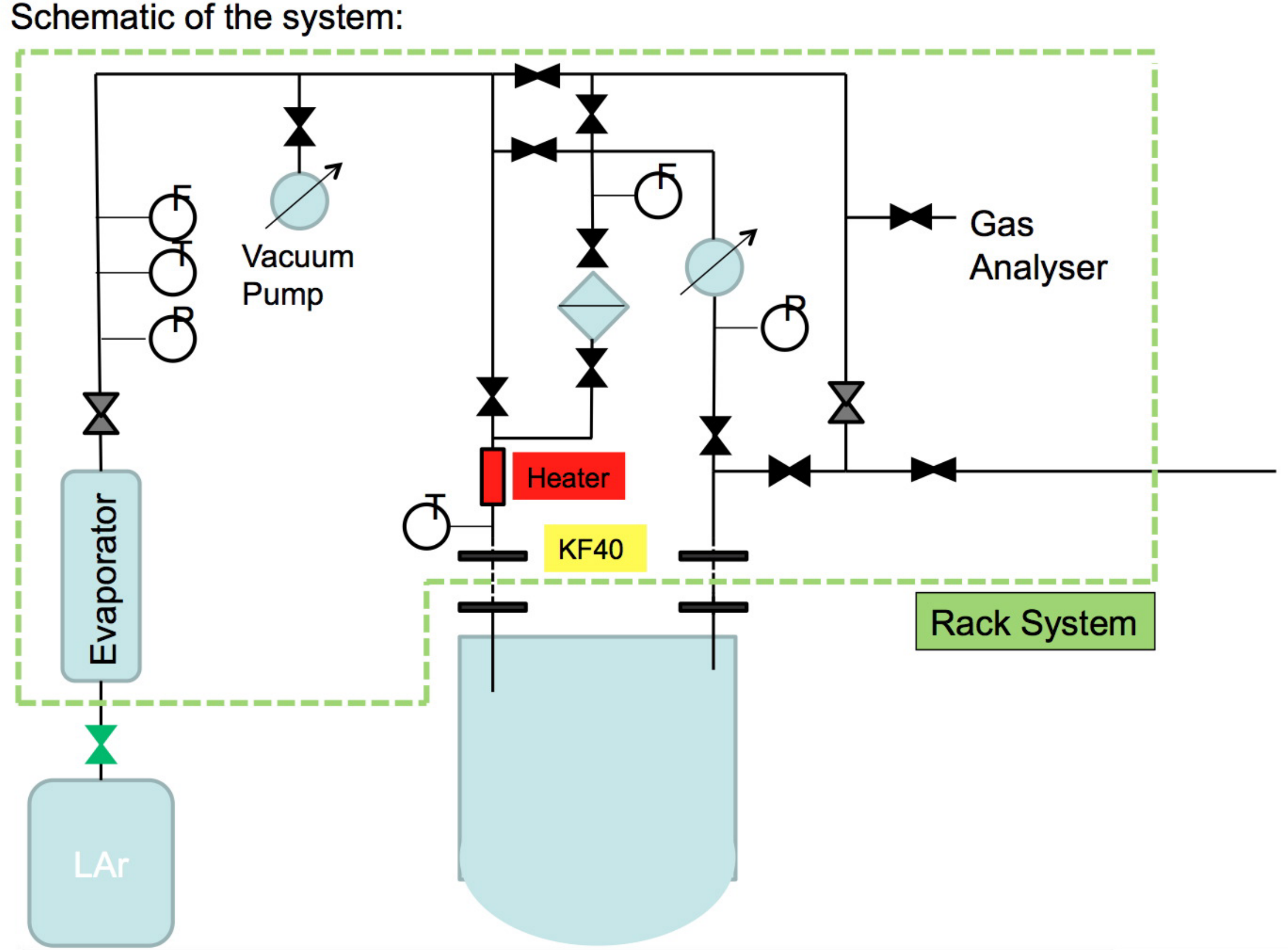}
  \caption{Schematics of the apparatus, completed with recirculation and gas analyzing systems.}
\label{f1}
\end{figure}

\section{Avoiding the ultra-high vacuum phase}
The vacuum phase 
covers at least three important separate functions in providing conditions for a low level of impurities in LAr:
\begin{enumerate}
\item it removes air from the vessel prior to cooling and filling with LAr;
\item it removes outgassing (water and other elements) from the detector materials and the vessel walls (baking
is also sometimes performed);
\item it verifies the integrity (tightness) of the system itself.
\end{enumerate}
In the present work, we focused on the item (i). Items (ii) and (iii) will be studied later. Concerning (i), 
it should be pointed out that
commercially available (high grade) bulk liquid argon, which is handled, transported and stored without using high vacuum 
vessels, has an oxygen content of about 1 ppm, which is an acceptable starting point for further 
purification in the liquid phase. Therefore it looks very reasonable to expect that at least the ``remove air''
function can be easily achieved without vacuum. 
Prior to our test, we therefore anticipated that the air in a vessel could be efficiently displaced by a flow of Ar gas, 
and we have experimentally tested this technique for reducing the air concentration using a 6 m$^3$ vessel. 

In our test, we first proved that oxygen can be effectively reduced from its air concentration (20\%) to less than 0.1\%
with oxygen detectors. However, in order to be sensitive to all potential contaminants that affect drift electron properties, 
we simultaneously instrumented our vessel with DUV photosensors (coated PMTs) to directly measure the 
argon scintillation light produced by embarked radioactive sources.
Indeed, several components are observed in the scintillation of gaseous argon. 
The mean lifetime of the slowest (triplet) component (about 3~$\mu$s) is very sensitive to traces of impurities and can be 
very effectively used to monitor the impurities~\cite{lumquench,KMthesis} that affect the drift properties of electrons.
See section~\ref{sec:impur}.

To our knowledge, this is the first time this sort of 
experiment is conducted in a vessel of this size at such level of contamination. 
A test of argon gas flushing was previously reported in Ref.~\cite{Jaskierny:2006sr}, however, the
vessel was smaller than in our case and only the oxygen content was monitored. In our work,
we therefore measured for the first time the level of {\it several}\ Êimpurities by a direct observation of the argon
scintillation light. 

More experimental work is ongoing, in particular to address the outgassing and leak checking aspects, 
and results are expected in the near future.

\begin{figure}[tb]
 \centering
   \includegraphics[width=0.6\textwidth]{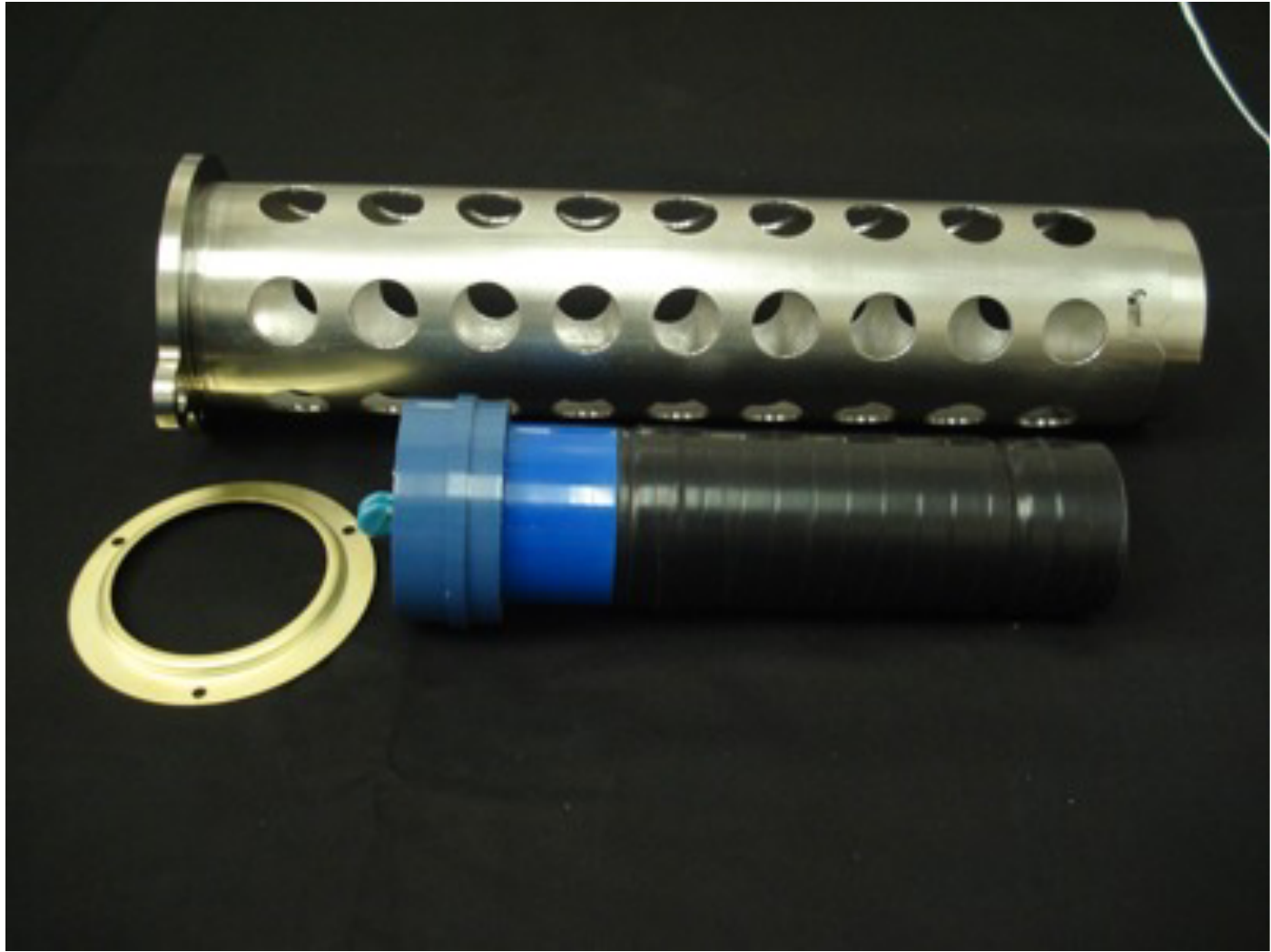}
  \caption{Photomultiplier tube (2'' ETL 9831KB) and corresponding stainless steel holder.}
\label{f3}
\end{figure}

\section{Experimental setup}

\subsection{Ar vessel}

A schematic of the apparatus, completed with a recirculation and gas analyzing systems, is shown in 
Fig.\ref{f1}.
A 6 m$^3$ vessel, previously used as a liquid argon dewar for testing of LAr/U calorimeters,  was 
refurbished and made available for this test.
The vessel is a double walled dewar, to allow for vacuum insulation in case of cryogenic operation. The 
walls of the inner vessel are stainless steel, while the top flange, 7 cm thick, is made out of aluminum, as 
the outer vessel. The diameter of the inner vessel is 200 cm and the total height of the
dewar is 235 cm. On the top flange there are 12 KF sealed nipples for feedthroughs. The top
flange is sealed with an o-ring. The tightness of the vessel has been measured through 
pressure decay leak testing, which allowed to determine a leak rate of 0.15 mbar 
lt / sec (which compares to a leak rate of $\sim$10$^{-6}$ mbar lt/sec characteristic of high 
vacuum vessels). 
This relatively large leak rate was deemed acceptable for our goals, since it corresponds - all things 
equal but a reversed pressure gradient - to a flow of 20 ppm of O$_2$ per hour; considering that the 
vessel is operated with an overpressure of $\sim$~100~mbar, the actual inflow of O$_2$ from air is likely 
to be rather negligible. The Ar gas transfer line provided a maximum flow of about 200 lt/min. The gas 
line was connected to a 35 mm diameter stainless steel tube which went all the way to the bottom of the 
vessel. 
The Ar gas was exhausted through a flow meter, which could measure a maximum flow of 100 lt/min. 

%

\begin{figure}[tb]
 \centering
   \includegraphics[width=0.9\textwidth]{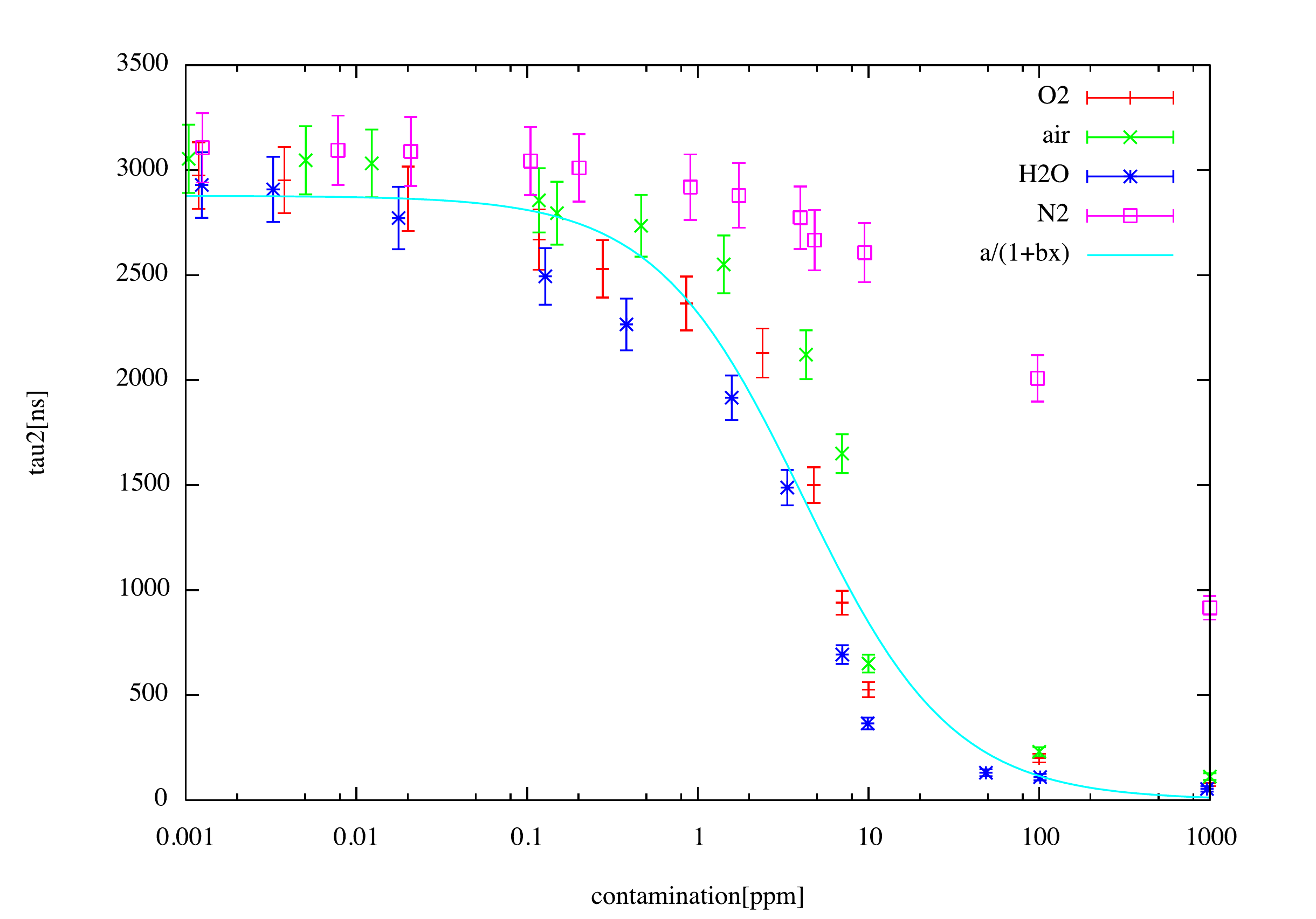}
  \caption{The dependence of $\tau _2$ from the concentration of different impurities in Ar gas. Data from  \cite{KMthesis}. A fit with a Birks' law-type function to the O$_2$ data is superimposed.}
\label{f2}
\end{figure}

\subsection{Detection of impurity traces with PMTs}\label{sec:impur}

During operation, the vessel was equipped with two percent level O$_2$ monitors\footnote{OXY-SEN 
model, from Alpha Omega Instruments.}, which are sensitive to a minimum concentration of 0.1\% O$_2$.
The oxygen sensors were left hanging from the top flange, one at a depth of 30~cm and the second one 
90~cm below.  

Three photomultiplier tubes (PMT) were deployed inside the vessel to monitor the DUV argon scintillation light, 
at a distance of 40, 110 and 180~cm from the top flange.
The PMTs are 2'' ETL 9831KB,  each housed in a cylindrical stainless steel holder with a 40 kBq $^
{241}$Am ($\alpha$ emitter) source  placed in front of the PMT, at a distance of 10 cm from the 
photocathode. There are large circular apertures on the sides of the holder, to allow for efficient gas 
circulation (Fig.\ref{f3}). Once inside the holder, a PMT sees a negligible amount of light from outside the 
holder itself, so that we could monitor the scintillation light of the argon at a given position. 
The inside of the PMT holder was lined with 3M foil specular reflector coated with 1 mg/cm$^2$ of 
tetra-phenylbutadiene (TPB), rendering the reflection form the surface 90\% diffuse~\cite{Boccone:2009kk}. In addition the PMT 
windows were coated with 0.05 mg/cm$^2$ TPB. TPB acts as a wavelength shifter for the DUV
scintillation light of argon to approximately 430 nm (blue), within the high quantum efficiency range 
of the PMT~\cite{Boccone:2009kk}.
The signal from each PMT was digitized at a sampling rate of 1 GS/s, using an Acqiris 
DP1400 digitizer. 

\begin{figure}
 \centering
   \includegraphics[width=0.6\textwidth]{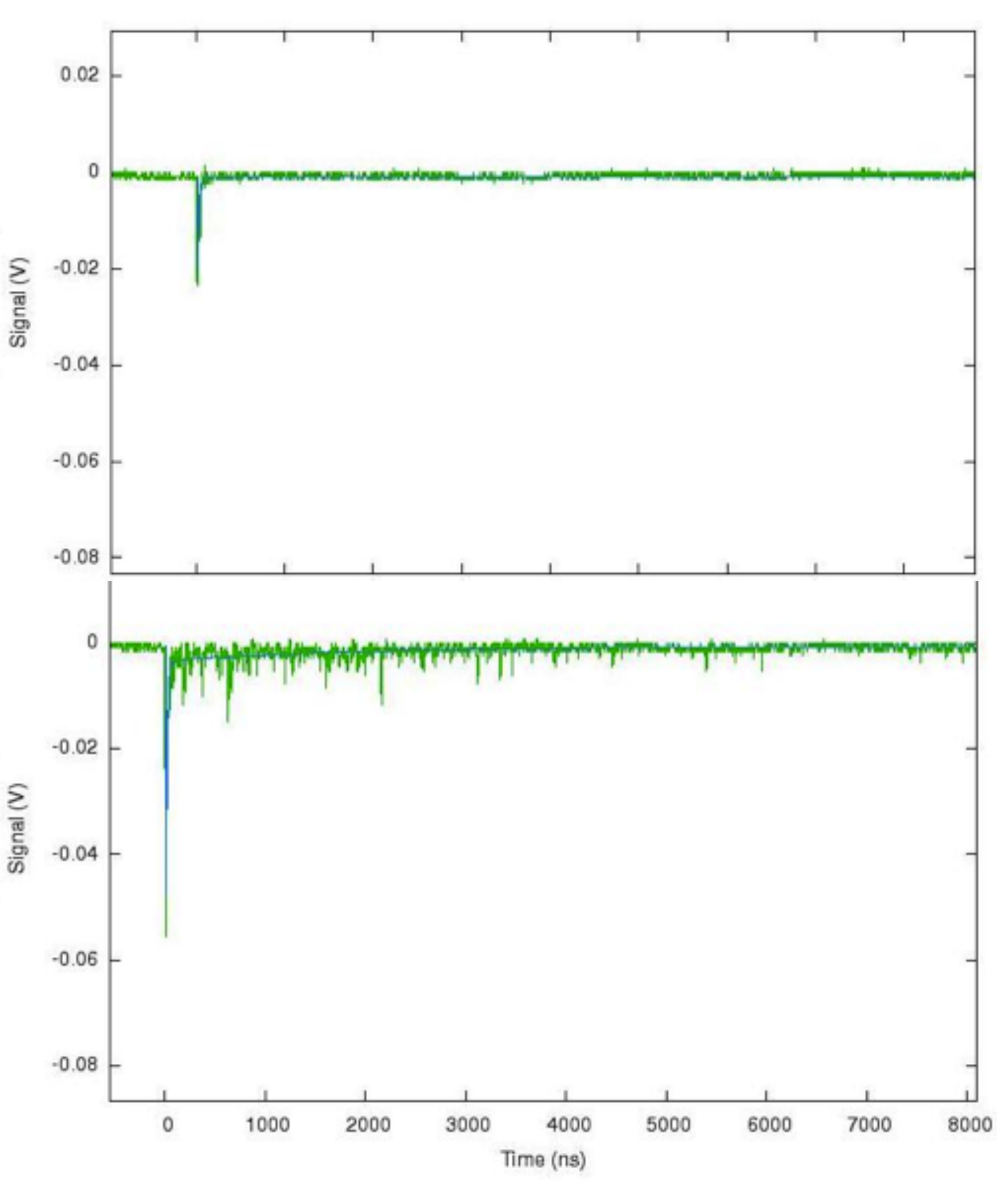}
  \caption{PMT traces for alpha particles from an Am source. {\it Top:} an event at the beginning of the 
  run, $\tau _2$ consistent with zero. {\it Bottom:} an event taken near the end of the run, with $\tau _2$ 
  exceeding 1200 ns. }
\label{f4}
\end{figure}

The scintillation light of pure argon gas has a fast and a slow component, with decay times $\tau_1$ 
shorter than 10 ns and  $\tau_2$ of about 3 $\mu$s \cite{Ar_scintillation_1, Ar_scintillation_2}. 
The fast component is rather independent of the purity of the Ar, while the decay time of the slow
component is sensitive to the concentration of impurities. In particular, very clean Ar shows a decay time 
exceeding 3 $\mu$s, while for O$_2$ concentration larger than 0.1\% the decay time is 100 ns or less.   
See for example Fig. \ref{f2} (data from \cite{KMthesis}), which is directly comparable to the 
measurements presented here since it was obtained with the same hardware and analysis procedure (in 
a smaller cell). 

The quenching effect of impurities on the slow component of the scintillation light can be roughly described by 
$$
\tau _2 \mathrm{[ns]} = \frac{\tau '}{1 + k \rho \mathrm{[ppm]}} 
$$
where $\tau '$ is the asymptotic value of $\tau _2$ for zero impurities, $\rho\mathrm{[ppm]}$ is the concentration of impurities in ppm and $k$ is a constant; this function is reminiscent of the Birks' law that describes the quenched light yield in 
scintillators for an increased $dE/dx$.
For $O _2$,  $\tau '$ = 2878 ns and $k$= 0.24.
This method is more sensitive for $\tau _2$ in the 500 to 2500 ns range (i.e. 0.5 to 10 ppm of O$_2$), 
while for smaller $\tau _2$ the variation is rather modest over a couple of orders of magnitude change in 
O$_2$ concentration.

For each measurement of the $\tau _2$ 10,000 events are acquired, a procedure 
which typically takes few seconds. Each waveform is fitted offline, to determine the value of $\tau _2$. 
The average value for the 10,000 events is taken as the measured $\tau _2$ at a given time. Traces 
taken at the beginning and at the end of the run are shown in Fig.\ref{f4}.


\section{Results}

\begin{figure}
 \centering
   \includegraphics[width=0.9\textwidth]{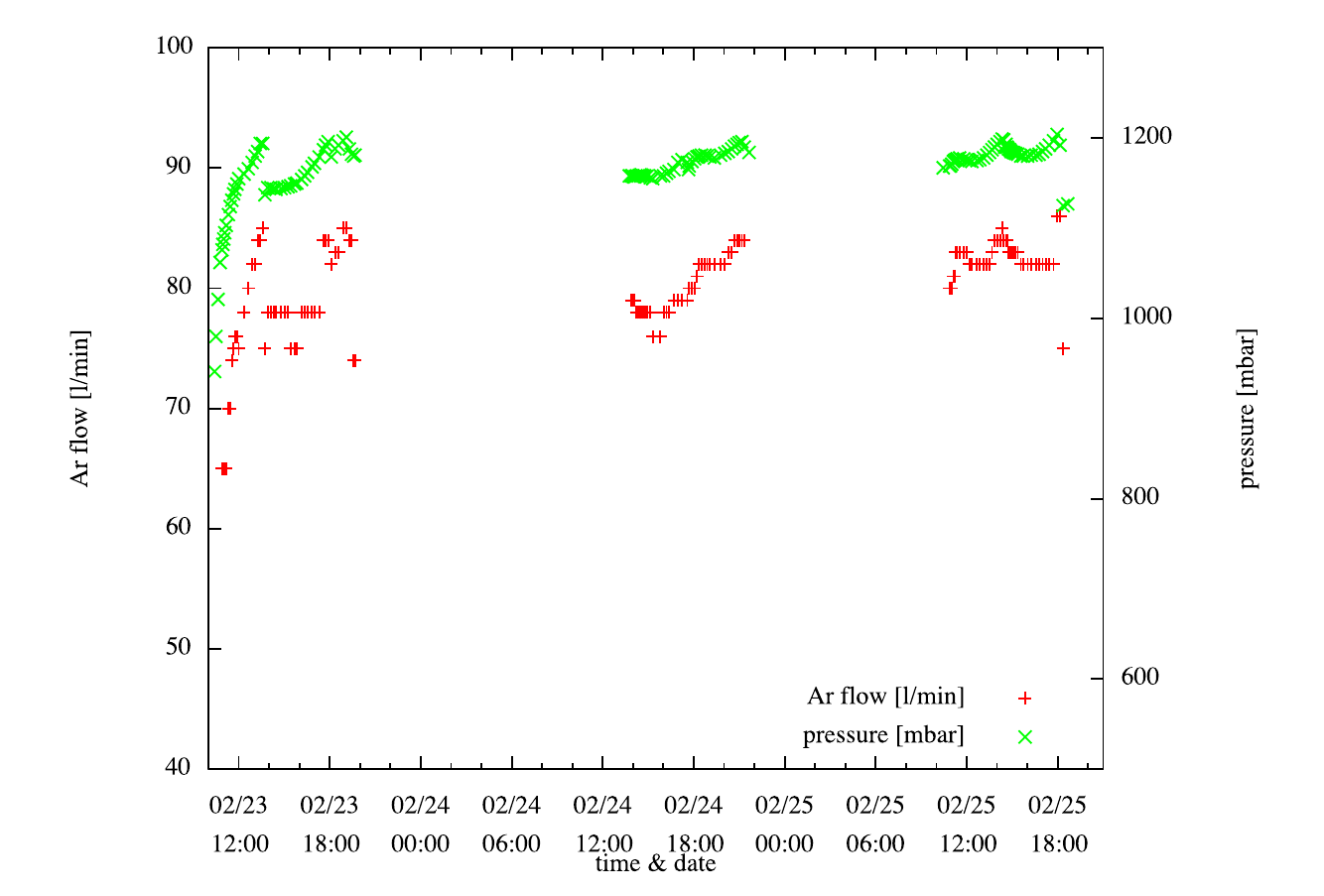}
  \caption{Ar gas flow and pressure in the vessel during Ar flushing.}
\label{f5}
\end{figure}

The 6~m$^3$ vessel was flushed with argon gas, not continuously, over more than a week. We started using a bottle 
(10 m$^3$ stp) of Ar57 (impurities less than 3 ppm) for a preliminary flushing which reduced the O$_2$ 
concentration from 20\% to 0.5\%, measured with one OXY-SEN sensor placed on the vessel exhaust. 
After 48 hrs without flushing any argon and the vessel with 150 mbar overpressure, the O$_2$ measured 
on the exhaust line actually improved, since purer argon had time to diffuse from the bottom to the top of the 
vessel.
A flange at the top was then opened to insert the two OXY-SEN sensors inside the ``clean" argon volume. 
Flushing with Ar57 was then restarted - as can be seen in Fig.\ref{f5}.
Part of the data from the OXY-SEN readings are collected in Tab.\ref{oxysen}; the two sensors at different 
depths show very clearly the pure Ar gas acting as a piston, removing the residual air starting from the 
bottom to the top. 
The Ar57 flow was approximately constant at 80 lt/min, limited by the flowmeter on the exhaust line; this 
is equivalent to one volume change per 75 min. 
The vessel was always  flushed with Ar57, except for the last 1.7 volume changes (starting at [02/25 14:00])
when Ar60 (impurities less than 1 ppm) was used, because the purification rate seemed to decrease, 
possibly due to $\sim$3 ppm of O$_2$  in the input gas.

\begin{figure}
 \centering
   \includegraphics[width=0.9\textwidth]{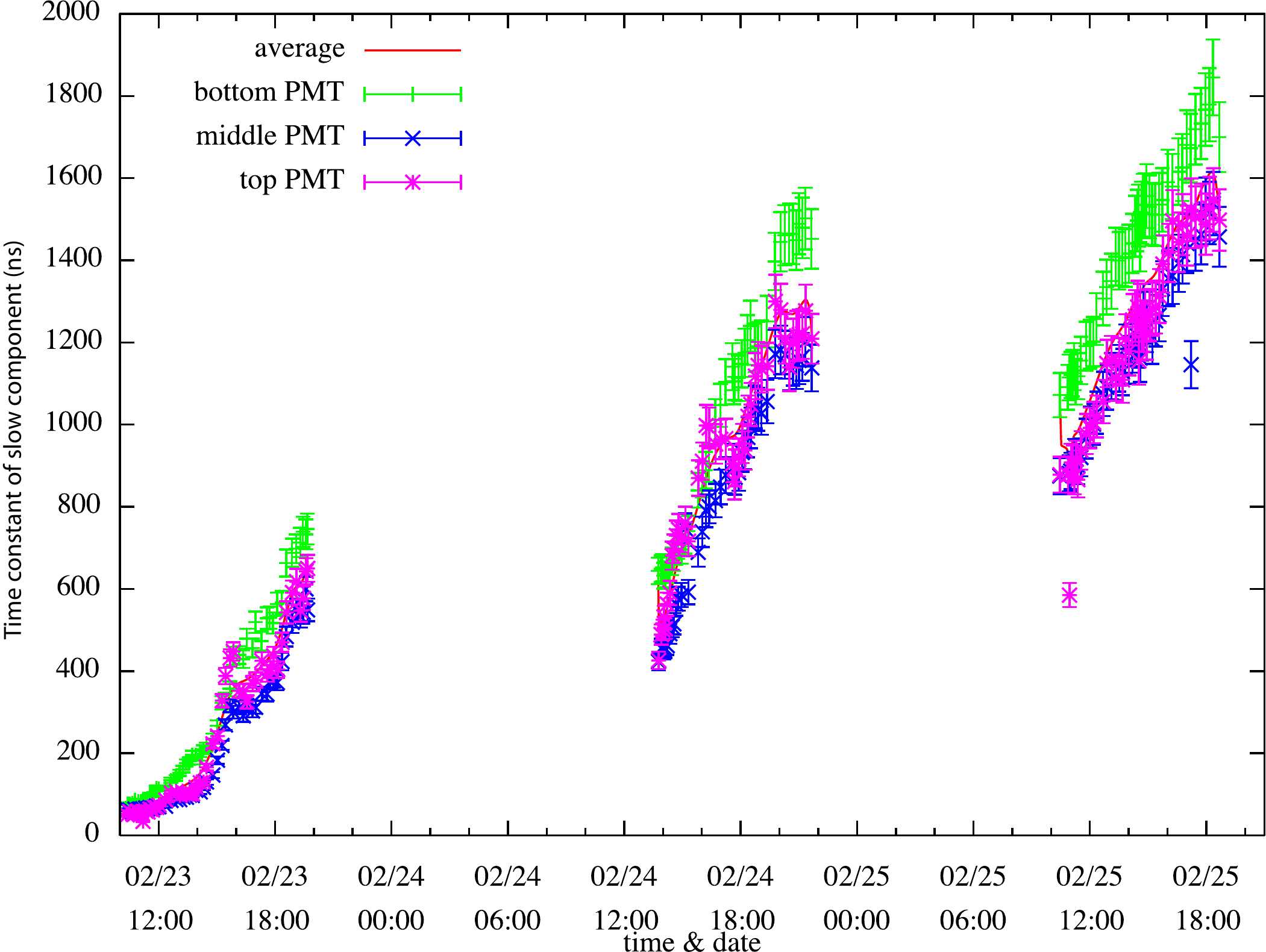}
  \caption{Fitted time constant of the slow component of the Ar scintillation light $vs.$time for three PMTs at 
  different depths. Data taken during Ar flushing.}
\label{f6}
\end{figure}

\begin{figure}
 \centering	
   \includegraphics[width=0.9\textwidth]{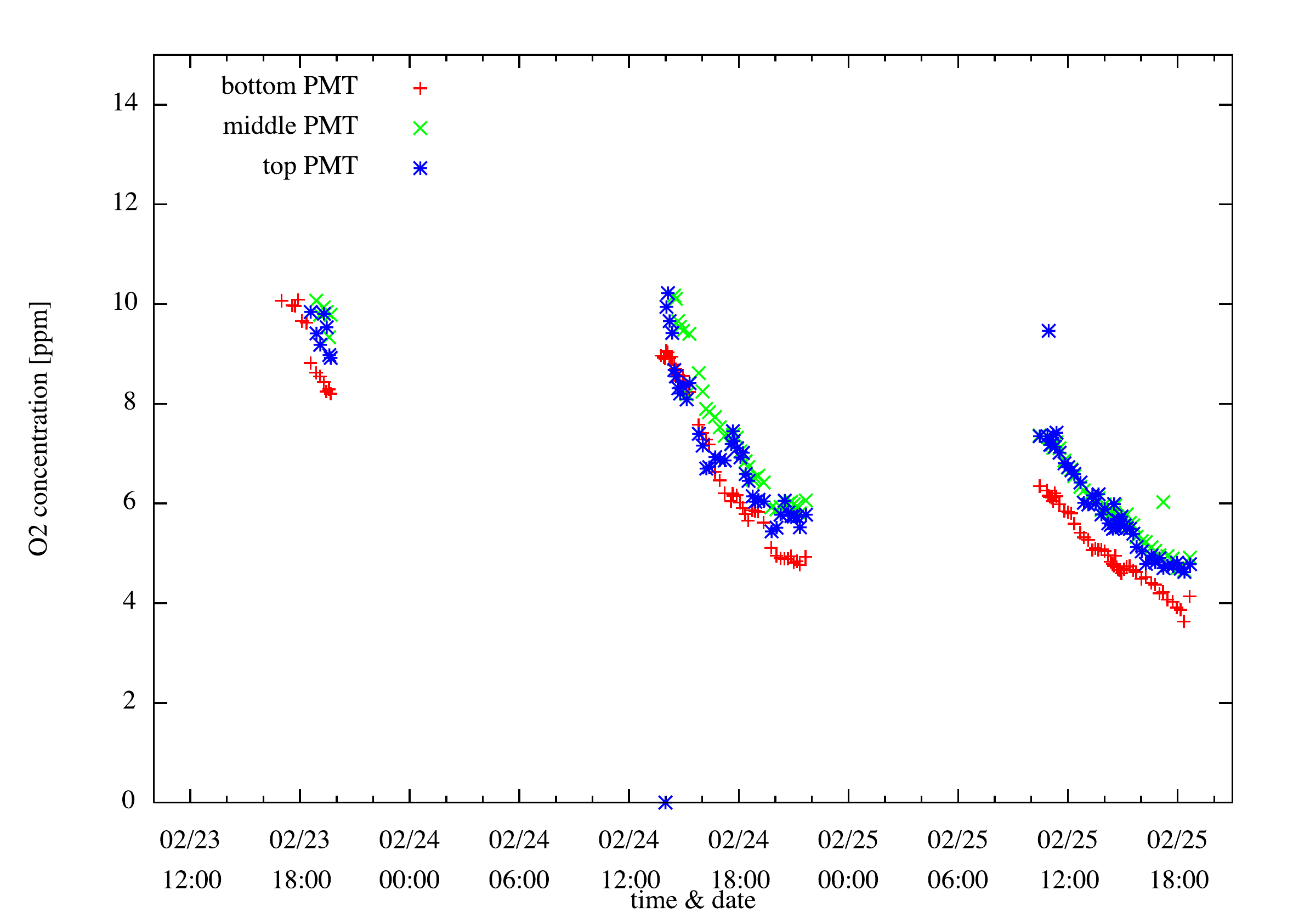}
  \caption{Same as Fig.\ref{f6}, converted in ppm of O$_2$ equivalent in argon.}
\label{f6b}
\end{figure}

Fig.\ref{f6} and Fig.\ref{f6b} summarize the main results versus acquisition time. 
When the sensitivity limit of the OXY-SEN sensors was reached (0.1\%), data have been taken 
continuously using the three PMTs. The decay time of the slow component of the Ar scintillation light ($
\tau _2$) improves continuously due to the removal of impurities. The PMT near the bottom of the 
vessel (and the Ar inlet) consistently shows the largest $\tau _2$, and shows the reduction in of 
contamination earlier than the other PMTs, as expected from the Ar gas acting as a piston.
The PMTs at the top and in the middle show a similar $\tau _2$, and the top one shows larger 
variations, correlated with sudden changes in the flow rate and pressure, which is measured at the top 
flange. 
The Ar flow near the top flange is expected to be quite turbulent and non-uniform, since the outlet is - in
practice - point-like when compared to the size of the flange itself. 
When the flushing stops, $\tau _2$ decreases by an amount roughly equivalent to the introduction of 
2ppm of O$_2$, possibly coming from outgassing and/or leaks to the outside at a rate of 0.15 ppm per 
hour.
The O$_2$ equivalent concentration at the end of the flushing procedure at the location of the top PMT is lower 
than 4 ppm, while the average of the 3 PMTs is slightly larger than 4 ppm. The O$_2$ equivalent concentration 
[ppm] $vs$ time is shown in Fig.\ref{f6b} for $\tau _2 > 500$ ns, i.e. the optimal range for this technique. 
The O$_2$ concentration drops of about a factor of two after 5 hrs of flushing, i.e. 24,000 lt of Ar gas stp 
or 4 volume changes. 

\begin{center}
\begin{table}[h]
\caption{\label{oxysen}O$_2$ concentrations as measured with OXY-SEN monitors (see text).}
\centering
\begin{tabular}{@{}*{7}{l}}
\br
Time\&Date 		& Exhaust 	& Top 		& Bottom \\
				& O$_2$ [\%]	& O$_2$ [\%]	& O$_2$ [\%] \\	
\mr
2010-02-19 15:30     & 19.1   		& -			& - \\
2010-02-19 16:18     & 9.1   		& -			& - \\
2010-02-19 16:28     & 6.9   		& -			& - \\
2010-02-19 16:39     & 5.2   		& -			& - \\
2010-02-19 16:48     & 4.2   		& -			& - \\
2010-02-19 16:57     & 3.5   		& -			& - \\
2010-02-19 17:06     & 3.3		& -			& - \\
2010-02-19 17:12     & 2.9		& -			& - \\    
2010-02-20 10:29     & 1.4 		& -			& - \\
2010-02-23 10:24 	& -			&	0.3     	&	0.3    \\
2010-02-23 10:30     & -			&	0.3    	&	0.3  	\\
2010-02-23 10:39     & -			&	0.3     	&	0.3	\\
2010-02-23 10:55	& -			&	0.3    	&	0.3	\\
2010-02-23 11:05     & - 			&	0.25    	&	0.3	\\
2010-02-23 11:20	& -			&	0.25		&	0.3	\\
2010-02-23 11:27	& -			&	0.2		&  	0.3   	\\
2010-02-23 12:01	& -			&	0.2		& 	0.3	\\
2010-02-23 12:21	& -			&	0.2		&     	0.25	\\
2010-02-23 12:38     & -			& 	0.2     	&	0.25 \\
2010-02-23 12:54	& -			&	0.2     	&	0.2 	\\
2010-02-23 13:35	& -			&	0.15    	&	0.2	\\
2010-02-23 13:44	& -     		&	0.1		&       0.2	\\
2010-02-23 16:09	& -			&	0.1		&    	0.15 \\
2010-02-23 16:21	& -     		&	0.1     	&	0.1	\\
2010-02-23 18:34	& -	     		&	0.1    	&	0.1 	\\

\br
\end{tabular}
\end{table}
\end{center}


%

%

\section{Conclusions}

The main results of this test is the purging of a 6 m$^3$ vessel from air to a residual 
contamination of few ppm of O$_2$ equivalent in pure Ar gas.  
The impurity concentration in argon has been measured down to the few ppm range using a simple 
system, based on the quenching of the Ar scintillation due to impurities.
The piston effect of argon displacing 
impurities has been seen both for percent level contamination and ppm level contamination. 
The final O$_2$ equivalent concentration, after ten volume changes, was $\sim$4 ppm, which corresponds to a 
reduction of the O$_2$ of about five orders of magnitude, faster than a simple exponential trend.
The experimental setup was remarkably simple, and the reduction in concentration of impurities in the 
vessel proved to be rather insensitive to interruptions in the flushing procedure. 
This is an experimental demonstration that impurities in argon can be reduced easily and 
efficiently to ppm level without vacuum pumping.
%


\ack{We thank the organizers of GLA2010 for inviting us to this excellent and very fruitful workshop.
This work was supported by ETH Z\"urich, the Swiss National Science Foundation (SNF) and the
University of Liverpool. We are grateful
to CERN for their hospitality where these tests were performed. }

\section*{References}

\end{document}